\documentclass[aps,pra,twocolumn,superscriptaddress,showpacs]{revtex4-1}

\usepackage{graphicx}
\usepackage[verbose=true]{epstopdf}
\usepackage{physics} 
\usepackage{hyphenat} 
\usepackage{hyperref}
\usepackage[dvipsnames]{xcolor}

\begin{document}

\newcommand{\etal}{\emph{et al.}}
\newcommand{\Li}{{}^{6}\textrm{Li}}
\newcommand{\LiTwoSOneHalf}{{2S_{1/2}}}
\newcommand{\UVtransition}{{2S_{1/2}\rightarrow 3P_{3/2}}}
\newcommand{\REDtransitionDTwo}{{2 \, S_{1/2}\rightarrow 2 \, P_{3/2}}}
\newcommand{\REDtransitionDOne}{2 \, S_{1/2}\rightarrow 2 \, P_{1/2}}
\newcommand{\GammaTwoP}{\Gamma_{2\mathrm{P}}}
\newcommand{\GammaThreeP}{\Gamma_{3\mathrm{P}}}
\newcommand{\Kfourty}{{}^{40}\textrm{K}}
\newcommand{\Kthirtynine}{{}^{39}\textrm{K}}
\newcommand{\Kfourtyone}{{}^{41}\textrm{K}}
\newcommand{\kB}{k_{\mathrm{B}}}
\newcommand{\Limp}{\textrm{Li}\left|3/2,3/2\right\rangle}
\newcommand{\LiOneOne}{\textrm{Li}\left|1/2,1/2\right\rangle}
\newcommand{\IsatTwoP}{I_{\mathrm{sat}}^{2\mathrm{P}}}
\newcommand{\IsatThreeP}{I_{\mathrm{sat}}^{3\mathrm{P}}} 
\newcommand{\UVMOT}{\textrm{UV\,MOT}} 
\newcommand{\LiSeven}{{}^{7}\textrm{Li}}
\newcommand{\LiDimerSix}{{}^{6}\textrm{Li}_2}
\newcommand{\LiDimerSeven}{{}^{7}\textrm{Li}_2}
\newcommand{\Isat}{I_{\mathrm{sat}}}
\newcommand{\Icool}{I_{\mathrm{cool}}}
\newcommand{\Irep}{I_{\mathrm{rep}}}
\newcommand{\LiGround}{{2\,S_{1/2}}}
\newcommand{\LiTwoPOne}{{2\,P_{1/2}}}
\newcommand{\LiThreePOne}{{3\,P_{1/2}}}
\newcommand{\LiTwoPThree}{{2\,P_{3/2}}}
\newcommand{\LiThreePThree}{{3\,P_{3/2}}}
\newcommand{\DarkState}{\psi_{\textrm{D}}}
\newcommand{\BrightState}{\psi_{\textrm{B}}}

\title{Comparison of an efficient implementation of gray molasses to narrow\hyp{}line cooling for the all\hyp{}optical production of a lithium quantum gas}

\author{C.~L.~Satter}
\affiliation{Centre for Quantum Technologies (CQT), 3 Science Drive 2, Singapore 117543}
\author{S.~Tan}
\affiliation{Centre for Quantum Technologies (CQT), 3 Science Drive 2, Singapore 117543}
\author{K.~Dieckmann}
\email[Electronic address:]{phydk@nus.edu.sg}
\affiliation{Centre for Quantum Technologies (CQT), 3 Science Drive 2, Singapore 117543}
\affiliation{Department of Physics, National University of Singapore, 2 Science Drive 3, Singapore 117542}

\date{\today}

\begin{abstract}
We present an efficient scheme to implement a gray optical molasses for sub\hyp{}Doppler cooling of $\Li$ atoms with minimum experimental overhead. To integrate the $D_1$ light for the gray molasses (GM) cooling into the same optical set\hyp{}up that is used for the $D_2$ light for a standard magneto\hyp{}optical trap (MOT), we rapidly switch the injection seeding of a slave laser between the $D_2$ and $D_1$ light sources. Switching times as short as $30\,\mu\textrm{s}$ can be achieved, inferred from monitor optical beat signals. The resulting low\hyp{}intensity molasses cools a sample of $N=9\times10^8$ atoms to about $60\,\mu\textrm{K}$. A maximum phase\hyp{}space density of $\rho=1.2\times10^{-5}$ is observed. On the same set\hyp{}up, the performance of the GM is compared to that of narrow\hyp{}line cooling in a UV MOT, following the procedure in \cite{Sebastian2014}. Further, we compare the production of a degenerate Fermi gas using both methods. Loading an optical dipole trap from the gray molasses yields a quantum degenerate sample with $3.3\times10^5$ atoms, while loading from the denser UV MOT yields $2.4\times10^6$ atoms. Where the highest atom numbers are not a priority this implementation of the gray molasses technique yields sufficiently large samples at a comparatively low technical effort. 

\end{abstract}

\pacs{37.10.De, 37.10.Gh, 67.85.Lm, 42.60.Fc}


\maketitle

\section{Introduction}
Ultracold quantum gases of neutral atoms have emerged as an ideal testing ground for quantum many\hyp{}body physics \cite{Bloch2008,Gross995}. Lithium is particularly suitable for this purpose due to the presence of a broad, magnetically tunable Feshbach resonance, which allows its two\hyp{}body interactions to be varied over a wide range. For the production of a quantum degenerate sample, a sufficiently high phase\hyp{}space density is necessary to initiate efficient evaporative cooling. For most species in the alkali\hyp{}metal group this can be achieved by standard Doppler and sub\hyp{}Doppler laser cooling on the $D_2$ transition. However for lithium (and potassium), standard laser cooling  to sub\hyp{}Doppler temperatures using the $D_2$ transition is inefficient. This is due to the fact that this transition is characterized by an unresolved hyperfine structure of the $\LiTwoPThree$ excited state \cite{Fort1998}. Cooling schemes for lithium that circumvent this problem are therefore of great interest, and efforts to improve these schemes form an active area of research.

So far, two robust methods are available that achieve cooling below the Doppler limit for $\Li$. In narrow\hyp{}line cooling of alkali atoms \cite{Duarte2011b,Sebastian2014,McKay2011a}, a magneto\hyp{}optical trap (MOT) is realized on the transition to the higher excited $\LiThreePThree$ state. This transition has a narrower linewidth and correspondingly a lower Doppler temperature. This technique was successfully used in our set\hyp{}up to create cold, dense samples of $\Li$ atoms \cite{Sebastian2014}, and to produce weakly\hyp{}interacting quantum\hyp{}degenerate Fermi gases and molecular BECs \cite{Gross2016}. 

The other cooling method is the gray optical molasses, which has been successfully employed in a series of experiments involving various alkali atoms \cite{Boiron1995,Boiron1996,RioFernandes2012,Grier2013,Nath2013,Salomon2013,Burchianti2014,Sievers2015,Colzi2016,Chen2016,Bruce2017,Rosi2018}. This method combines velocity\hyp{}selective coherent population trapping (VSCPT) into dark states with a friction force arising from a Sisyphus mechanism \cite{Shahriar1993,Grynberg1994,Weidemueller1994}. The gray molasses operates on the open atomic transitions $F \rightarrow F$ and $F \rightarrow F-1$, for which $\Lambda$\hyp{}shaped energy level structures are available that allow the occurrence of dark states. Therefore, the gray molasses is commonly applied on the $D_1$ transition. In many previous experiments a separate laser system was used to generate the light at the $D_1$ transition. Often, the $D_1$ and $D_2$ frequency beams are superimposed as a final step before being guided to the experiment using an optical fiber. In this way, $D_1$ gray molasses can be performed subsequently to standard $D_2$ loading of a MOT.  

For the all\hyp{}optical production of quantum degenerate gases it can be beneficial to enhance the number density of the atomic sample for improved loading into an optical dipole trap (ODT). The gray molasses method does not provide a confinement force during the cooling, and hence does not actively increase the density. In this case, an improved ODT loading efficiency can only be realized with additional experimental effort, e.g.~by increasing the spatial overlap between the cooled cloud and the ODT (see e.g.~\cite{Burchianti2014}), or by adapting the pre\hyp{}cooling stages to achieve higher initial densities before applying the GM (see e.g.~\cite{Salomon2013}).



\begin{figure*}
	\centering
	\includegraphics[width=\textwidth]{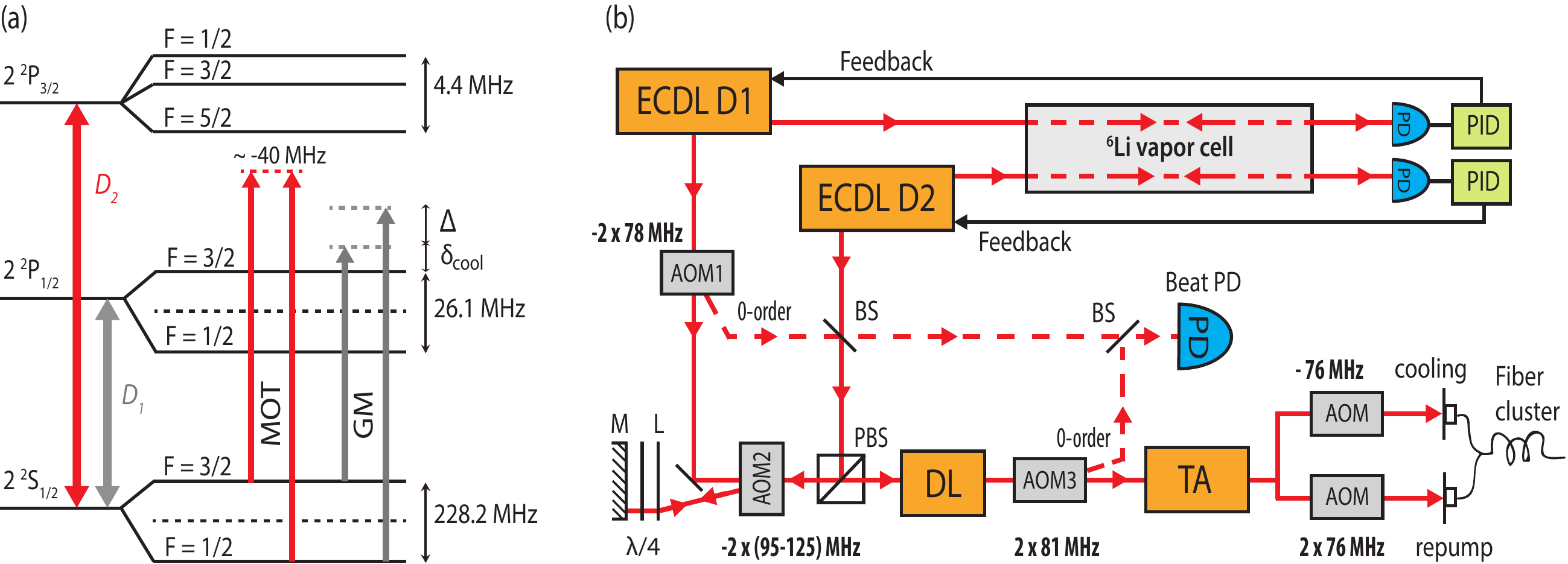}
	\caption[The integrated set\hyp{}up for the production of gray molasses light]{(a) Level scheme showing the $D_1$ and $D_2$ transitions for $\Li$. Arrows indicate the transitions used for the cooling and repumping frequencies in the experiment. (b) The integrated set\hyp{}up for the production of MOT and gray molasses cooling light. The $D_2$ and $D_1$ frequency beams seed the slave diode laser (DL) in turn, and from that point onwards follow the same path to the experiment. A beat system is used to monitor the seeding of the DL at each instant in the experimental sequence. Double pass AOM configurations are indicated in the schematics by a factor of two written in front of the radio frequency. For AOM2, it is drawn out in detail, to demonstrate the introduction of the $D_1$ light into the set-up.}
	\label{fig:IntegratedSetup}
\end{figure*}

In this article we present an efficient implementation of simple gray molasses (GM) cooling on the $D_1$ transition for $\Li$ that minimizes experimental overhead. This implementation leads to the production of a degenerate Fermi gas without additional methods to enhance the ODT loading. 
For our implementation of the gray molasses scheme only one external cavity diode laser (ECDL) and an acousto\hyp{}optic modulator (AOM) were added to a pre\hyp{}existing laser system for cooling on the $D_2$ transition. Here, it is demonstrated how the $D_1$ light is integrated by rapidly switching the injection seeding of a common slave diode laser (DL) between $D_2$ and $D_1$ light. Seeding of the DL is monitored through the use of a beat between seeding light and output light. In the following, the integrated laser system and beat set\hyp{}up are discussed, followed by a characterization of the performance of the GM. On the same set\hyp{}up we present a comparison of the GM performance to that of narrow\hyp{}line cooling in a two\hyp{}stage MOT, following the procedure in \cite{Sebastian2014}. We believe that such a direct comparison is of value for the design of many new or existing experiments with lithium or potassium where different laser cooling strategies are considered. To further extend the comparison and to assess the overall performance of our approach to GM cooling, we present the production of a strongly interacting, degenerate Fermi gas using both schemes. 

\section{Experimental Approach}
The energy level structure of atomic $\Li$ is shown in Fig.~\ref{fig:IntegratedSetup}(a). We perform gray molasses cooling on the blue side of the $\ket{F=3/2}\rightarrow \ket{F'=3/2}$ transition, with repumping light on the $\ket{F=1/2}\rightarrow \ket{F'=3/2}$ transition. The light at the $D_2$ and $D_1$ frequencies is produced in the set\hyp{}up illustrated in Fig.~\ref{fig:IntegratedSetup}(b). Both the $D_1$ and $D_2$ ECDLs are frequency stabilized to the respective atomic transitions using Doppler\hyp{}free frequency modulation spectroscopy in one vapor cell. For the $D_2$ MOT the output of the $D_2$ ECDL is frequency\hyp{}shifted by AOM2 and seeds a slave DL. This allows for frequency tuning around the MOT transitions without variation of the optical power. The optical power of the slave DL is sufficient to seed a tapered amplifier (TA), which amplifies the $25\,\textrm{mW}$ seed beam in a single pass to up to $300\,\textrm{mW}$. The output beam is split into two beams and further shifted by AOMs to address the cooling and repumping transitions. A polarization\hyp{}maintaining optical fiber cluster (Evanescent Optics Inc.) with two input and six output ports is used to combine the cooling and repumping beams and transport the light to the experiment. At the output, the MOT beams are expanded to a $1/e^2$ radius of $0.9\,\textrm{cm}$. In total, up to $15\,\textrm{mW}$ (or $4.5\,\Isat$, where $\Isat=2.54\,\textrm{mW/cm}^2$ is the saturation intensity of the $D_2$ line) is available in each of the six beams, with an intensity ratio between cooling and repumping light that can be regulated. 

For GM cooling the $D_1$ light enters the set\hyp{}up from the ECDL and AOM1 through AOM2. AOM2 is set up in double\hyp{}pass configuration, and the $D_1$ light enters along the unused zero\hyp{}order direction only on the second pass. During the MOT cooling stage the $D_1$ beam path is shuttered, but at the end of this stage AOM2 is quickly switched off to interrupt the path for the $D_2$ light and to allow the $D_1$ light to pass through AOM2 unrefracted and hence parallel to the direction of the $D_2$ light. From there, it seeds the slave diode laser, and follows the same beam path as the $D_2$ light through the TA to the experiment. AOM1 is used to enable fast switching of the $D_1$ light. With this arrangement the $D_1$ and $D_2$ light can be rapidly swapped, and no further alignment is needed. Furthermore, this means that the gray molasses beams have a beam diameter identical to the MOT beams and have the same $\sigma_{+}$/$\sigma_{-}$ polarization, which is suitable for gray molasses cooling \cite{RioFernandes2012}. For the purpose of controlling the parameters relevant for GM cooling, AOM1 is used to obtain the required detuning $\delta_{\textrm{cool}}$ to the blue side of the $D_1$ cooling transition. The two AOMs before the fiber cluster control the intensities of the cooling and repumping light, respectively $\Icool$ and $\Irep$, and the detuning $\delta_{\textrm{rep}}$ of the repumping light.

Since the frequencies of the $D_1$ and $D_2$ atomic transitions are separated by only $10.1\,\textrm{GHz}$, the slave diode laser can be injection seeded by either source ECDL. However, the different light frequencies match the laser diode cavity modes at different values of the DL operating current. To quickly switch between the two, as is done when progressing from the MOT cooling stage to the gray molasses, a current modulation pulse is sent to the DL. 
Because gray molasses cooling is only applied for a few milliseconds, a single modulation current value is typically sufficient. Temperature fluctuations due to the sudden change in injection current are negligible for timescales up to at least $10\,\textrm{ms}$, as was observed by varying the duration of the gray molasses phase.

\begin{figure}[t]
	\centering
	\includegraphics[width=0.48\textwidth]{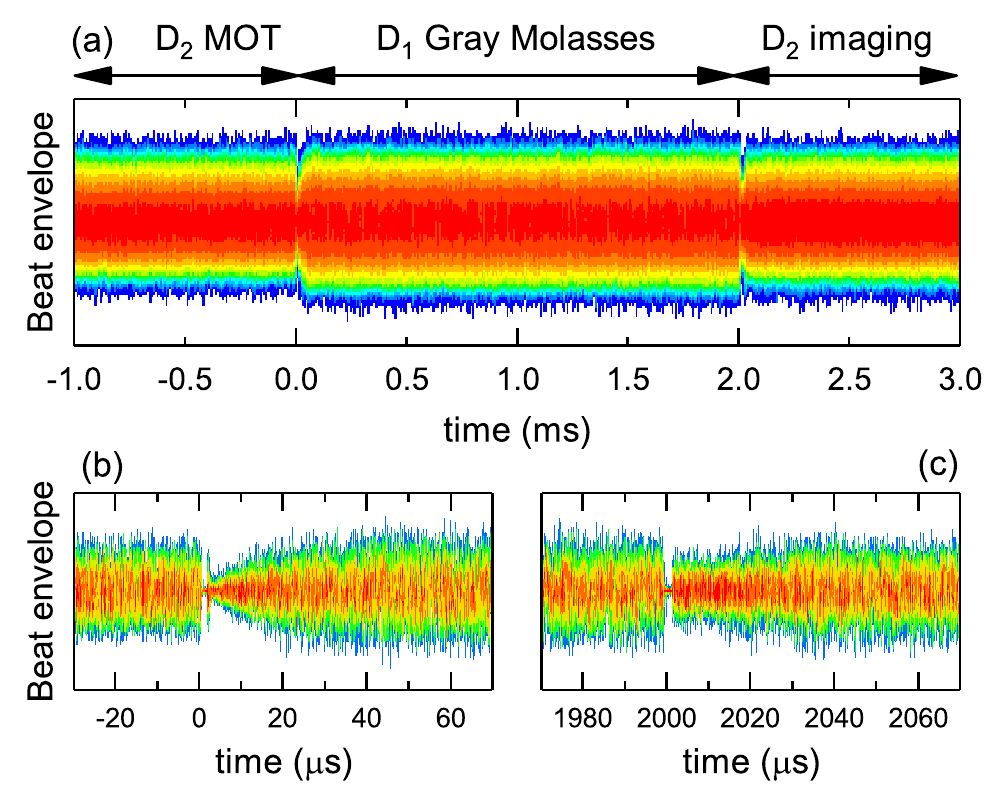}
	\caption{The envelope of the $D_2$ and $D_1$ beats as a function of time during the experimental sequence. Here, the sample density is displayed, as the beat signals are rapidly oscillating ($>100\,\textrm{MHz}$) on the observed timescales. Warmer colors represent higher sample density. At $t=0$, the $D_2$ light was switched off, and the path was opened for the $D_1$ light to seed the DL. After $2.0\,\textrm{ms}$ of gray molasses cooling, the source for seeding the DL is switched back to the $D_2$ ECDL for absorption imaging. Panel (a) shows the beat PD signal over the entire beat sequence, while panels (b) and (c) show the signal at the switch-over points with higher time resolution.}
	\label{fig:BeatEnvelope}
\end{figure}

As a convenient method to monitor whether the DL is truly following the $D_1$ frequency for the entire duration of gray molasses cooling and how quickly it does so, a beat system was set up. The zero\hyp{}order beams from AOM1 after the $D_1$ ECDL and AOM3 after the DL are overlapped onto a photodiode (PD), as is indicated by the dashed beam paths in Fig.~\ref{fig:IntegratedSetup}(b). When the DL follows the optical frequency dictated by the $D_1$ ECDL, a beat appears at a frequency of $156\,\textrm{MHz}$, which is twice the frequency applied to AOM1. In the experiment we monitor the beat by observing its envelope. If no injection locking is established the beat ceases and the observed amplitude vanishes. Using this method, the response of the slave DL can be monitored during the experiment and, if necessary, be acted upon accordingly, e.g. by adjusting the modulation current to the DL. An example of a typical signal as observed during the experimental sequence is shown in Fig.~\ref{fig:BeatEnvelope}(a). When the trapping and cooling cycle is completed, the atomic ensemble is imaged using laser light at the $D_2$ frequency. For this purpose, the source for seeding the slave laser needs to be switched back to the $D_2$ ECDL. To monitor that this happens correctly and quickly enough, another beat is produced, this time between the $D_2$ ECDL and DL outputs. In practice we see, as shown in Fig.~\ref{fig:BeatEnvelope}(b) and (c), that the DL can switch very quickly between the two light sources, with a gap of only $\approx 30\,\mu\textrm{s}$ in between during which it is not properly seeded.

\begin{figure}[tbp]
	\centering
	\includegraphics[width=0.45\textwidth]{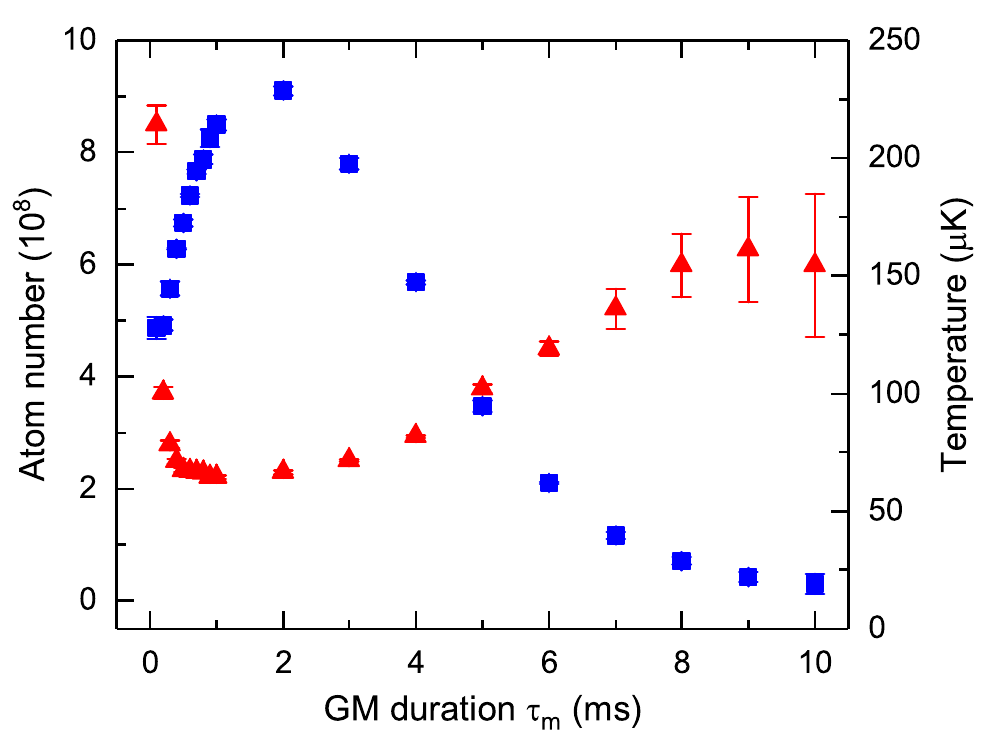}
	\caption[Temporal evolution of the gray molasses]{Temporal evolution of the gray molasses. The number of captured atoms $N$ (blue squares) and temperature $T$ (red triangles) are plotted as a function of the molasses duration $\tau_{\textrm{m}}$. The data correspond to the gray molasses at a magnetic bias field of $-0.9\,\textrm{G}$.}
	\label{fig:GMDuration}
\end{figure}

\section{The gray molasses}
Before GM cooling is applied, our experimental sequence starts with a standard 3D MOT on the $D_2$ transition. After $15\,\textrm{s}$ of loading from a Zeeman\hyp{}slowed atomic beam, the MOT is compressed by simultaneously decreasing the detuning and light intensity, and increasing the gradient of the magnetic quadrupole field. At the end of the compression phase $1.9\times10^9$ atoms remain in the compressed MOT (cMOT), at a peak density of $n_0=3.5\times10^{10}\,\textrm{cm}^{-3}$. From time\hyp{}of\hyp{}flight absorption images we infer a typical temperature of $800\,\mu\textrm{K}$.

\begin{figure*}[htbp]
	\centering
	\includegraphics[width=\textwidth]{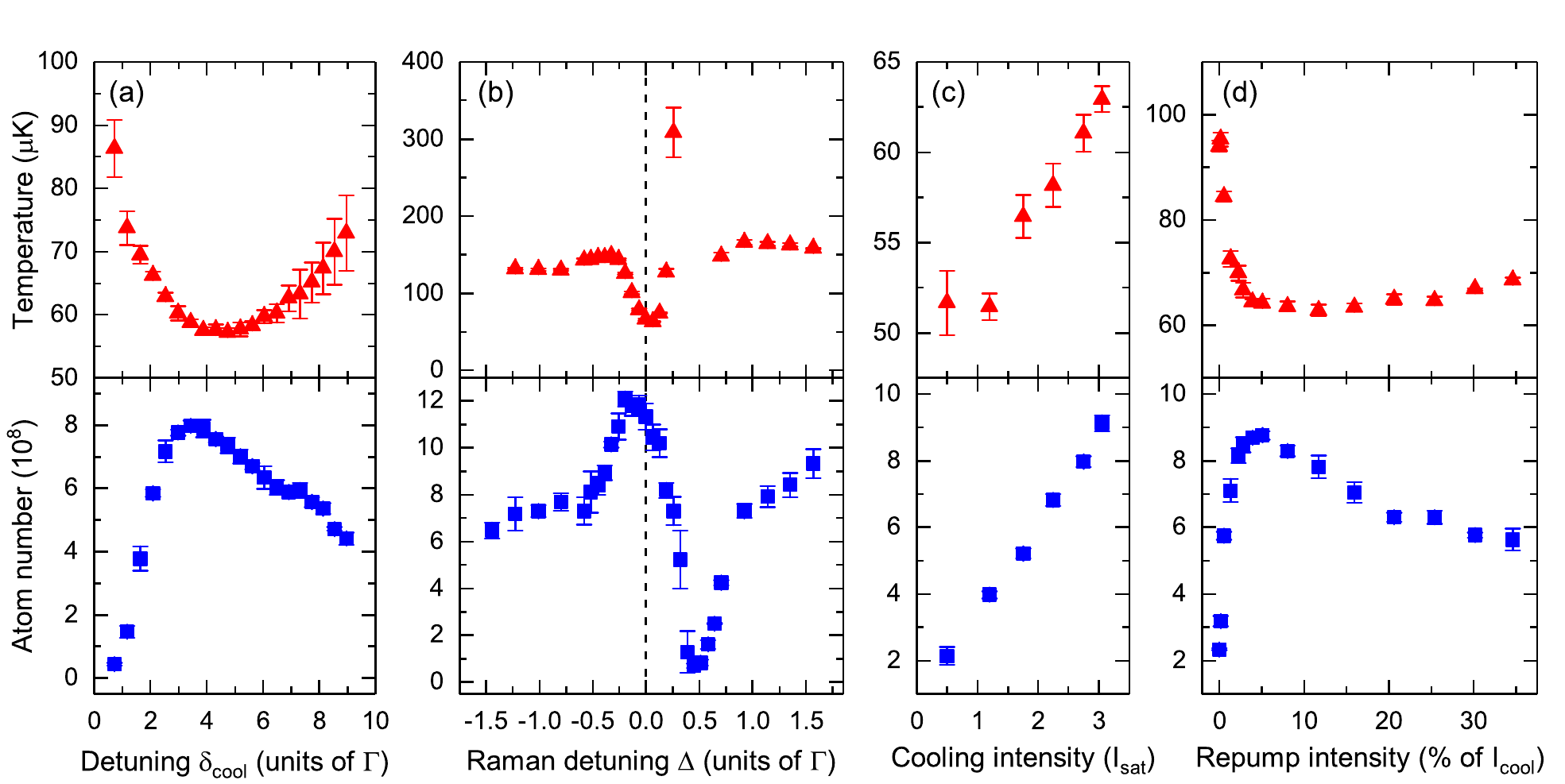}
	\caption{Number of cooled atoms (blue squares) and temperature (red triangles) as a function of (a) the detuning $\delta_{\textrm{cool}}$, (b) the Raman detuning $\Delta$ between repumping and cooling light, (c) the intensity of the cooling light, and (d) the intensity of the repumping light. For (b), around $\Delta=0.5\,\Gamma$ the temperature data are not presented, as a partially heated cloud was observed.}
	\label{fig:GMMaster}
\end{figure*}

In preparation for the gray molasses cooling stage, the $D_2$ light is extinguished and the magnetic quadrupole field is switched off. In our chamber this induces a magnetic eddy field to which the gray molasses cooling proved sensitive. In order to allow the field to decay, the $D_1$ light is turned on by the AOMs in front of the fiber cluster only after a delay of $0.5\,\textrm{ms}$. The earth field as well as the remaining eddy field were compensated by applying a constant bias field during GM cooling. In the direction along the axis of the MOT coils, which required the most significant compensation, a bias field of $-0.9\,\textrm{G}$ was applied. This value was chosen to yield the lowest temperatures at short molasses durations. We measured a quadratic increase in temperature of $\Delta{T}\propto 210\,\mu\textrm{K}/\textrm{G}^2$ when deviating from this optimum.

\subsection{Characterization}
We first studied the evolution of the gray molasses over time. In Fig.~\ref{fig:GMDuration} the atom number $N$ and temperature $T$ are plotted against the duration $\tau_{\textrm{m}}$ of the $D_1$ molasses pulse. For these measurements an overall blue detuning of $\delta_{\textrm{cool}}=3.2\,\Gamma$ and an intensity of $\Icool=3.1\,\Isat$ were used. Here, $\Gamma=2\pi\times5.87\,\textrm{MHz}$ is the natural linewidth and $\Isat$ refers to the saturation intensity of the $D_2$ transition. The Raman detuning between cooling and repumping light was kept constant at $\Delta=\delta_{\textrm{rep}}-\delta_{\textrm{cool}}=-0.045\,\Gamma$ and $\Irep$ was kept at 5\% of $\Icool$. The sample temperature was observed to decrease rapidly during the first $0.5\,\textrm{ms}$, and reached a minimum of $T\simeq64\,\mu\textrm{K}$. At $\tau_{\textrm{m}}=2\,\textrm{ms}$, a maximum of $50\%$ of cMOT atoms was captured by the molasses and cooled. After $\tau_{\textrm{m}}=3\,\textrm{ms}$ however, the temperature started to rise and the atom number dropped. This is attributed to the fact that the magnetic bias field applied compensates well for the eddy fields that are present during the first few milliseconds of the cooling stage, but is not optimal for longer durations, when the eddy fields have further decayed.

In the following, we describe the effect of different parameters on the performance of the gray molasses for a $2\,\textrm{ms}$ GM pulse duration, as summarized in Fig.~\ref{fig:GMMaster}. Varying the overall laser detuning the capture efficiency and cloud temperature reach optima at a blue detuning of $\delta_{\textrm{cool}}=4\,\Gamma$ with respect to the $D_1$ transition, as can be seen from Fig.~\ref{fig:GMMaster}(a). Upon increasing the detuning further, the captured atom number decreases, and cooling is less effective. This is expected, since for optical molasses the capture velocity is proportional to $1/{\delta^2}$. 

To show the sensitivity of the GM cooling to the Raman detuning $\Delta$ we pre\hyp{}cooled the cloud at $\Delta=-0.045\,\Gamma$ for $2\,\textrm{ms}$ and applied a $0.25\,\textrm{ms}$ pulse at a variable Raman detuning. As can be seen from Fig.~\ref{fig:GMMaster}(b), cooling is optimal at exact Raman resonance, while the number of atoms captured displays a clear maximum for a small, red Raman detuning $\Delta=-0.2\,\Gamma$. For $\Delta$ chosen slightly blue of the Raman condition, a strong heating of the cloud occurs, accompanied by a sharp decrease in the number of cooled atoms. The observed Fano profile formed by the dependence of atom number and temperature on $\Delta$ has become a signature of gray molasses cooling  (see e.g. \cite{Grier2013,Nath2013,Salomon2013}). The strong contribution of the repumper and the enhanced cooling near the Raman condition point to the existence of long\hyp{}lived coherences between the two hyperfine manifolds $\ket{\LiGround,F=3/2}$ and $\ket{\LiGround,F=1/2}$. In the presence of the repumper, a second $\Lambda$\hyp{}type three\hyp{}level system is formed, with new inter\hyp{}manifold dark states and bright states \cite{Grier2013}.

The effect of the intensity of the GM cooling light on the atoms is shown in Fig.~\ref{fig:GMMaster}(c). During this measurement, the ratio $I_{\textrm{rep}}/I_{\textrm{cool}}$ was kept constant. We observe that, as the cooling intensity is increased, the number of atoms captured by the molasses increases linearly. This is in agreement with the dependence of the capture velocity on the intensity, as $v_{\textrm{cap}}\sim\Gamma_{\textrm{p}}/k$, where $\Gamma_{\textrm{p}} \propto I/{\delta^2}$ is the optical pumping rate. Since the trend is not yet seen to saturate at $I_{\textrm{cool}}=3\Isat$, we expect that a larger number of atoms can be captured into the molasses by using higher\hyp{}intensity $D_1$ beams. The temperature of the molasses is observed to increase linearly with the intensity. This is expected, as the equilibrium temperature pertaining to a Sisyphus cycle scales with the light shift as $T \propto I/{\delta}$ \cite{RioFernandes2012}.

The intensity of the repumper $I_{\textrm{rep}}$ with respect to the cooling intensity $I_{\textrm{cool}}$ also turned out to play an important role. In Fig. \ref{fig:GMMaster}(d), the captured atom fraction and temperature are plotted against $I_{\textrm{rep}}$ in units of percentage of $I_{\textrm{cool}}$. The captured atom fraction shows a clear maximum at $5.5\%$, which corresponds to a repumping intensity of $I_{\textrm{rep}}=0.12\,\Isat$.  The decrease in atom number for increased repumping power can be attributed to increased photon reabsorption of the scattered light on the cooling transition, as the optical density of the cloud with respect to the cooling transition increases with the repumper intensity \cite{Landini2011}. 

In our experiment, the coldest temperature after the gray molasses was observed when the Raman detuning was tuned to exact $\Delta=0$, for the full GM duration of $2\,\textrm{ms}$. Then, a temperature of $41(2)\,\mu\textrm{K}$ can be obtained, at the cost of the captured atom number being reduced to half of the maximum value. Presumably, even lower temperatures can be reached by using a low cooling light intensity. Several gray molasses experiments (e.g. \cite{RioFernandes2012,Salomon2013,Sievers2015,Colzi2016}) take advantage of the relationship between cooling intensity and temperature to implement a two\hyp{}stage GM. A high intensity is used to maximize the capture efficiency, after which the intensity is ramped down to obtain lower temperatures. In our set\hyp{}up, this method was observed not to yield a temperature improvement. In this we believe we are limited by the low available beam intensity and long\hyp{}lived eddy fields, which prevent us from using the GM for a duration long enough to implement intensity ramps. 

We calculate the phase\hyp{}space density from $\rho = n_0 \Lambda_T^3$, where $\Lambda_T=\frac{h}{\sqrt{2\pi m k_\mathrm{B} T}}$ is the thermal de Broglie wavelength and $k_\textrm{B}$ is the Boltzmann constant. The highest phase\hyp{}space density is attained at $\Delta=-0.05\,\Gamma$ and $2\,\textrm{ms}$ GM duration. Up to $N=9\times10^8$ cooled atoms are detected in the gray molasses, at a temperature of around $60\,\mu\textrm{K}$. This corresponds to a captured fraction of $50\%$ from the cMOT. The peak density was inferred from \textit{in situ} imaging. As noted earlier,  there are no confinement or compression forces present during the gray molasses, and hence the peak density does not increase during GM cooling. Since a fraction of the atoms remains uncooled by the molasses, the peak density decreases with respect to the cMOT to $n_0=2\times10^{10}\,{\textrm{cm}}^{-3}$. This results in a maximum phase\hyp{}space density of $\rho=1.2\times10^{-5}$. 

\subsection{Comparison with narrow\hyp{}line MOT}

In the following we compare the results from the GM cooling to the narrow\hyp{}line MOT previously described in \cite{Sebastian2014}, which employs the $\UVtransition$ at an ultraviolet (UV) wavelength. For this purpose the UV MOT is loaded after an initial MOT stage on the $D_2$ transition, which is almost identical to the one used in the GM experiment. However, due to the different magnetic field requirements of the UV MOT, the bias fields during both the red MOT and UV MOT stage are allowed to take on more beneficial values as compared to the situation where the GM scheme is used. This leads to a lower red cMOT temperature of $T=480\,\mu\textrm{K}$ with a total number of atoms of $N=1.4\times10^9$ captured. 

The parameters for the UV MOT are similar to the ones previously described. For repumping light the red $D_2$ transition is used. The UV MOT is loaded for $1.5\,\textrm{ms}$ at a detuning of $-8\,\GammaThreeP$ and magnetic field gradient of $3.4 \,\textrm{G/cm}$ to optimize the number of atoms captured from the red cMOT. In another $1.5\,\textrm{ms}$ a compression is carried out, during which the detuning is ramped up to $-2.9\,\GammaThreeP$ and the field gradient is increased to $12 \,\textrm{G/cm}$. After compression is completed, the UV MOT reaches a transient maximum in phase\hyp{}space density after another $16\,\textrm{ms}$ of hold time. 

In the current measurements, $1.3\times10^8$ atoms were observed to be present in the UV MOT after the compression stage, at a temperature of $T=70\,\mu\textrm{K}$. This leads to a peak density of $n_0=1.2\times10^{11}\,\textrm{cm}^{-3}$ and phase\hyp{}space density of $\rho=7\times10^{-5}$ after the UV cMOT.

When it comes to the peak density of the atomic cloud, narrow\hyp{}line cooling in the UV MOT has the distinct advantage that it operates in the presence of a magnetic quadrupole field. This means that the atoms can be spatially confined during cooling, and can be compressed by increasing the magnetic field gradient. Moreover, in our set\hyp{}up the UV MOT cooling beams are independent of the beams for the $D_2$ MOT, providing us with the flexibility to choose a smaller $1/e^2$ beam waist of $ 4.3\,\textrm{mm}$. The result is that peak density and phase\hyp{}space density are approximately a factor of six higher as compared to the values attained for the case of gray molasses cooling. In the following we investigate how this difference affects the transfer efficiency into the small volume and limited depth of an optical dipole trap and evaporative cooling into quantum degeneracy.

\section{Production of a strongly interacting Fermi gas}
In this section we investigate the suitability of our implementation of the gray molasses for the all\hyp{}optical production of a lithium quantum gas. We directly compare this to the results obtained with narrow\hyp{}line cooling. For this purpose we follow the same procedure as previously described in detail in \cite{Gross2016}. In this scheme the pre\hyp{}cooled atoms are transferred into a two\hyp{}beam, small\hyp{}angle, crossed optical dipole trap (ODT), transported from the MOT vacuum chamber into a glass science cell, and are evaporatively cooled. For the purpose of evaporative cooling of the fermions we make use of the broad Feshbach resonance in the spin mixture of the two energetically lowest Zeeman levels $\ket{F=1/2,m_F=\pm1/2}$ (commonly referred to as states $\ket{1}$ and $\ket{2}$). Our crossed optical dipole trap has a transverse beam waist of $ 66\,\mu\textrm{m}$. For the measurements presented here we start evaporation with an initial optical power of $22.5\,\textrm{W}$ per beam. This corresponds to a trap depth of approximately of $ k_\textrm{B}\times 330\,\mu \textrm{K} $ and axial and radial trap frequencies of $ \omega_{z} = 2\pi \times 64(9) \, \textrm{Hz} $ and $ \bar{\omega}_{r} = \sqrt{ \omega_x \omega_y }  = 2 \pi \times 2.81(3) \, \textrm{kHz} $, respectively.

\subsection{Loading the ODT from the gray molasses}
For the transfer of the atoms into the ODT from the gray molasses, the optical power is ramped up within $ 0.5\,\textrm{ms} $ to the full trap depth at the start of the GM phase. Both the GM and ODT beams are on for $ 1.5\,\textrm{ms} $, during which time the gray molasses reaches its highest density, and coldest conditions. We observed heating and atom loss from the trap for longer cooling periods as expected from the measurements of the gray molasses presented in Fig.~\ref{fig:GMDuration}. The GM is then switched off by extinguishing the repumping light in advance of the cooling light, to optically pump the atoms into the $\ket{F=1/2}$ hyperfine ground state manifold. This creates an incoherent balanced mixture of the $\ket{1}$ and $\ket{2}$ states.

\begin{figure}[t]
	\includegraphics[width=0.45\textwidth]{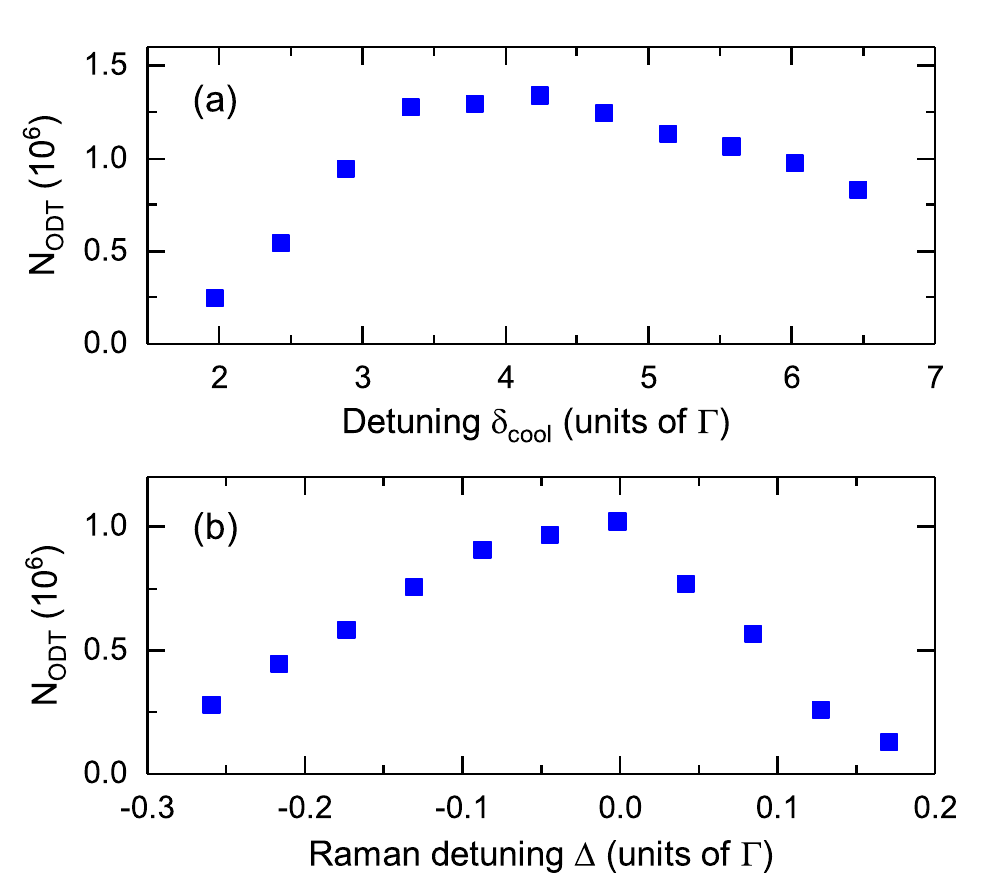}
	\caption{(Color online) The number of atoms captured from the GM into the ODT as a function of (a) the overall detuning $\delta_{\textrm{cool}}$, and (b) the Raman detuning $\Delta$.}
	\label{fig:ODTLoading}
\end{figure}

While the operating wavelength of $1070\,\textrm{nm}$ for the ODT was chosen to be near the magic wavelength \cite{Safronova2012} for the UV MOT transition, this is not the case for the gray molasses. Hence, it is worth investigating the possibility that the light shift induced by the ODT could affect the efficacy of GM cooling by shifting the global detuning of the cooling light. We investigated the effect by varying the detuning of the gray molasses light during the combined cooling and optical trapping stage. Fig.~\ref{fig:ODTLoading} shows the atom number yield. As can be seen from Fig.~\ref{fig:ODTLoading}(a), no significant shift was observed, if $\delta_\textrm{cool}$ was varied. This is consistent with the experimental determination by Burchianti $\etal$ \cite{Burchianti2014} of a light\hyp{}shift coefficient of $+6.3(7)\,\mathrm{MHz/(MW/cm^2)}$ for the case of a $\lambda=1073\,\textrm{nm}$ trapping laser wavelength. From this we expect for our ODT power a frequency shift of $2\,\textrm{MHz}$, less than a natural linewidth. The critical sensitivity of the atom number yield to the Raman detuning $\Delta$ was observed in the ODT as well, as shown in Fig.~\ref{fig:ODTLoading}(b). This demonstrates the dependence of the loading efficiency on the phase\hyp{}space density produced by the GM.

Immediately after loading, we observe that $N = 1.35 \times 10^6 $ atoms are transferred to the optical dipole trap. The temperature is $ T = 63 \,\mu\textrm{K} $. Following the loading, we allow for $1\,$s of plain evaporation. For the purpose of thermalization, a magnetic bias field of approximately $ 200 \,\textrm{G} $ is applied. This tunes the s\hyp{}wave scattering length between the two states to $a_{12} = -220\, a_0 $ \cite{Zurn2013}, where $a_0$ is the Bohr radius. After plain evaporation the ODT typically contains a total number of atoms of $ N = 1.1 \times 10^6 $ at a temperature of $ T = 43 \,\mu\textrm{K} $, corresponding to $T/T_\textrm{F} = 7.3$, where $T_\textrm{F}$ is the Fermi temperature. This corresponds to an initial truncation parameter of $\eta=7.7$ before forced evaporation is applied. 

\subsection{Loading the ODT from the UV MOT}
We perform a similar loading sequence when the ODT is loaded from the UV MOT. For the transfer of the atoms into the crossed ODT, the optical power is linearly ramped up within $3\,\textrm{ms}$ to the full trap depth during the compressed UV MOT phase. A combined loading and cooling duration of $11\,$ms is used, in order to allow for the peak density to reach its transient maximum value. Afterwards, the UV MOT is switched off and hyperfine pumping and plain evaporation are done in the same manner as described in the previous section for the GM case. Afterwards, the ODT typically contains a total number of atoms of $N=7.5\times10^6$ at a temperature of $45\,\mu\textrm{K}$ corresponding to $T/T_\textrm{F}=4.2$ and a truncation parameter of $\eta=7.3$ before initiating forced evaporation. The atom number found is about a factor of seven higher as compared to the GM case, which is in good agreement with the ratio in phase space density produced by the two cooling methods.
 
\subsection{Evaporation to degeneracy}
To facilitate creating a strongly interacting quantum gas in the BEC\hyp{}BCS crossover regime, forced evaporation is performed near the broad $832 \,\textrm{G}$ Feshbach resonance. Here, we apply a magnetic bias field of $ 940 \, \textrm{G} $ to evaporate on the BCS side above the resonance. At this field the $s$\hyp{}wave scattering length is $ a_{12} = -5100\, a_{0} $ \cite{Zurn2013}, establishing rapid thermalization during the cooling. The trap depth is reduced according to the theoretical curve 
$U_0 \left(t\right) = U_0 \left( 0\right) \left( 1 - t/\tau_\textrm{u} \right)^{2\left(\eta^{\prime}-3\right) / \left( \eta^{\prime} - 6 \right)}$, which was derived by Luo $\etal$ \cite{Luo2006} for a Fermi gas mixture with a unitarity\hyp{}limited s\hyp{}wave scattering cross section. 
$\eta^{\prime}$ is related to the truncation parameter by $\eta^{\prime}=\eta+\left(\eta-5\right)/\left(\eta-4\right)$. 

\begin{figure}
	\includegraphics[width=0.45\textwidth]{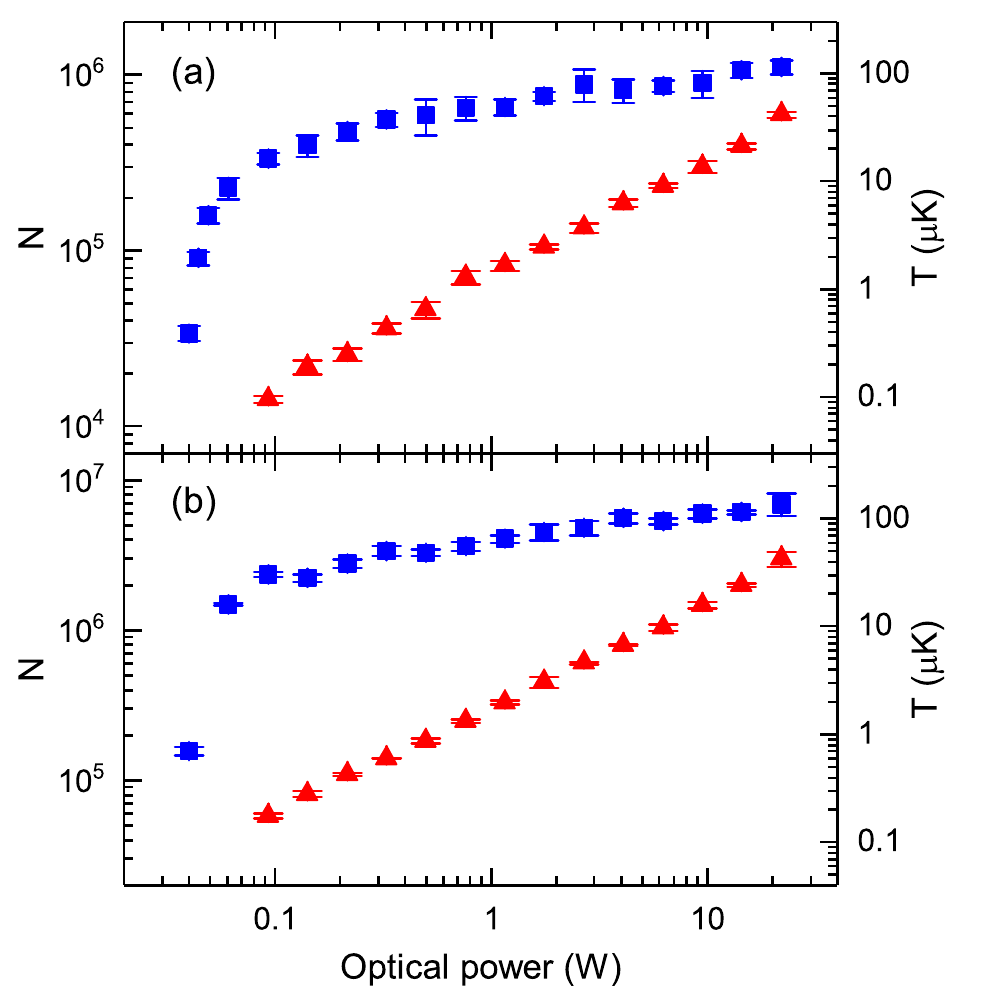}
	\caption{(Color online) Total number of atoms $N$ (blue squares) and temperature $T$ (red filled circles) at various points during forced evaporative cooling performed at $940 \, \textrm{G}$. The x\hyp{}axis displays the optical power per beam of the ODT at which the evaporation ramp was halted. Panel (a) the ODT was loaded from the gray molasses, time constant: $10\,\textrm{s}$ (b) the ODT was loaded from the UV MOT, time constant: $3\,\textrm{s}$. 
	}
	\label{fig:EvaporationUVGM}
\end{figure}

For the GM case we use a time constant of $\tau_\textrm{u} = 10\,s$, which was experimentally optimized for maximum atom number. For this time constant it takes about $8.5\,$s to lower the optical power to about $40\,$mW per beam, at which point atoms are increasingly spilled from the trap. To characterize the evaporation we probe the atomic cloud at various intermediate optical powers during the evaporation ramp. The images are fitted to a Fermi profile, from which we extract atom number, temperature, and $ T / T_\textrm{F} $. The results are shown in Fig.~\ref{fig:EvaporationUVGM}(a). We observe efficient cooling of the cloud down to a laser power of $P=93 \, \textrm{mW} / \textrm{beam} $, where $3.3(3)\times10^{5}$ atoms remain trapped at a temperature of $96(7) \, \textrm{nK}$, corresponding to $T/T_{\text{F}}=0.30(2)$. The atom number rapidly decreases with lower optical powers, which indicates that atoms are spilling from the trap, as the trap depth is lowered below the Fermi energy.

For the case of loading from the UV MOT the time constant for evaporation ramp is allowed to be smaller, due to a higher initial density of atoms in the ODT. Here, a time constant of $\tau_\textrm{u}=3\,\textrm{s}$ was used corresponding to a duration of $2.6\,\textrm{s}$. The results are presented in Fig.~\ref{fig:EvaporationUVGM}(b). The temperature as a function of trap depth follows a nearly identical curve as in Fig.~\ref{fig:EvaporationUVGM}(a), especially at higher trap depths. Evaporation is seen to be efficient down to optical powers of $93\,$mW, where $2.4(1)\times10^6$ atoms remain trapped at a temperature of $175(8)\,\textrm{nK}$, corresponding to $T/T_\textrm{F}=0.28(1)$. As expected the atom number in the quantum degenerate regime is higher than for the GM case. The relative factor of about seven found at the end of the plain evaporation stage remains nearly constant throughout the forced evaporation ramp. The higher atom number in this case might lead to systematic deviations in the temperature measurement near degeneracy due collisional hydrodynamic expansion. However, further investigation of the thermometry is beyond the scope of this work.

\section{Conclusion}
In summary, we have demonstrated a simple experimental implementation of a gray molasses and utilized it to produce a quantum degenerate gas of lithium. This was achieved by rapid switching of the injection seeding of a diode laser between the $D_2$ and $D_1$ light needed for loading into a standard MOT and gray molasses cooling, respectively. We demonstrated how a beat between seeding and output light of the diode laser can be used as a convenient method to monitor the correct operation of the gray molasses experiment. The gray molasses was characterized and compared with the UV MOT where narrow\hyp{}line cooling is applied. While both methods reach similar sub\hyp{}Doppler temperatures, the low\hyp{}intensity gray molasses yields a larger cloud containing a higher number of atoms, whereas the UV MOT produces a smaller but denser cloud with a higher phase\hyp{}space density. This resulted in a higher transfer efficiency of the atoms into a crossed optical dipole trap from the UV MOT than from the GM. After evaporative cooling this translated into a proportionally larger sample size at quantum degeneracy. We therefore find that with minimal experimental effort, the gray molasses allows us to obtain a quantum degenerate cloud with on the order of a few $10^5$ atoms. For many purposes, such as the modern experiments on quantum gas microscopy, this atom number is already sufficient. Where a larger degenerate cloud is needed, and certain  additional experimental effort is acceptable, the UV MOT offers a significant improvement of the sample size. 

\subsection*{Acknowledgments}
This research is supported by the National Research Foundation, Prime Ministers Office, Singapore and the Ministry of Education, Singapore under the Research Centres of Excellence program.


\end{document}